\PassOptionsToPackage{unicode}{hyperref}
\PassOptionsToPackage{hyphens}{url}
\documentclass[
  a4paper,
]{article}
\usepackage{amsmath,amssymb}
\usepackage{lmodern}
\usepackage{iftex}
\ifPDFTeX
  \usepackage[T1]{fontenc}
  \usepackage[utf8]{inputenc}
  \usepackage{textcomp} 
\else 
  \usepackage{unicode-math}
  \defaultfontfeatures{Scale=MatchLowercase}
  \defaultfontfeatures[\rmfamily]{Ligatures=TeX,Scale=1}
\fi
\IfFileExists{upquote.sty}{\usepackage{upquote}}{}
\IfFileExists{microtype.sty}{
  \usepackage[]{microtype}
  \UseMicrotypeSet[protrusion]{basicmath} 
}{}
\makeatletter
\@ifundefined{KOMAClassName}{
  \IfFileExists{parskip.sty}{%
    \usepackage{parskip}
  }{
    \setlength{\parindent}{0pt}
    \setlength{\parskip}{6pt plus 2pt minus 1pt}}
}{
  \KOMAoptions{parskip=half}}
\makeatother
\usepackage{xcolor}
\usepackage{graphicx}
\makeatletter
\def\maxwidth{\ifdim\Gin@nat@width>\linewidth\linewidth\else\Gin@nat@width\fi}
\def\maxheight{\ifdim\Gin@nat@height>\textheight\textheight\else\Gin@nat@height\fi}
\makeatother
\setkeys{Gin}{width=\maxwidth,height=\maxheight,keepaspectratio}
\makeatletter
\def\fps@figure{htbp}
\makeatother
\newlength{\cslhangindent}
\newlength{\csllabelwidth}
\newlength{\cslentryspacingunit} 
\newenvironment{CSLReferences}[2] 
 {
  \setlength{\parindent}{0pt}
  \ifodd #1
  \let\oldpar\par
  \def\par{\hangindent=\cslhangindent\oldpar}
  \fi
  \setlength{\parskip}{#2\cslentryspacingunit}
 }%
 {}
\usepackage{calc}

\usepackage{float}
\usepackage{changepage}
\ifLuaTeX
  \usepackage{selnolig}  
\fi
\IfFileExists{bookmark.sty}{\usepackage{bookmark}}{\usepackage{hyperref}}
\IfFileExists{xurl.sty}{\usepackage{xurl}}{} 
\DeclareRobustCommand{\href}[2]{#2\footnote{\url{#1}}}
\hypersetup{
  pdftitle={Automatically Finding and Categorizing Replication Studies},
  pdfauthor={Authored by Bob de Ruiter},
  hidelinks,
  pdfcreator={LaTeX via pandoc}}

\title{Automatically Finding and Categorizing Replication Studies}
\author{Authored by Bob de Ruiter}
\date{}

\begin{document}
\maketitle
\begin{abstract}
In many fields of experimental science, papers that failed to replicate
continue to be cited as a result of the poor discoverability of
replication studies. As a first step to creating a system that
automatically finds replication studies for a given paper, 334
replication studies and 344 replicated studies were collected.
Replication studies could be identified in the dataset based on text
content at a higher rate than chance (AUROC = 0.886). Additionally,
successful replication studies could be distinguished from failed
replication studies at a higher rate than chance (AUROC = 0.664).
\end{abstract}

\thispagestyle{empty}
\newpage

\hypertarget{introduction}{%
\section{Introduction}\label{introduction}}

Replication studies are attempts to validate previous research by
repeating the research methods. In many fields of experimental science,
an alarming number of replication studies has results or conclusions
deviating from the original research. Perhaps best known are large-scale
collaborations such as the Reproducibility Project, which failed to
replicate 62 out of 97 psychology papers (Nosek et al. 2018), even
though all original papers were widely-cited and published in esteemed
journals.

Most replication efforts, however, are not part of such a collaboration.
Because replication studies do not have a single defining
characteristic, finding all published replications of a paper can prove
to be difficult, be it automatically or manually.

If all replications of a given study could be automatically identified,
checking the integrity of a paper's bibliography would become much
easier. Another benefit would be that citation databases and academic
search engines such as Google Scholar could include the replications of
each paper on the result page, increasing their visibility.

In this paper, exploratory research is performed to investigate whether
any of this is possible. As far as I am aware, no research of this kind
has ever been done before. Because of this, I restrict my scope to
looking at whether replication studies can be distinguished from
replicated studies and whether failed replications can be distinguished
from successful replications.

Since replication studies typically cite the papers they replicate,
replication searches can be limited to papers that cite the original
paper. However, replication studies only account for a fraction of the
citations of most papers. As such, additional filtering is needed. One
approach is to only query for documents that contain variants of the
words ``replication'', ``reproduction'', ``reanalysis'' and
``reinvestigation'', but this does not even return half of all
replication studies. Another problem with a simple rule-based system is
that replication studies also cite papers they do not replicate.

Instead of constructing a rule-based system, a supervised text
classification model could be trained to identify replication studies.
This, however, requires a labeled dataset. No such dataset can be found
in one place. Fortunately, the University of Göttingen's ReplicationWiki
(Höffler 2017) has a
\href{http://replication.uni-goettingen.de/wiki/index.php/Category:Replication}{collection
of structured data concerning a hundreds of replication attempts spread
over as many pages}. The initial collection was put together by
students. According to a disclaimer on the website, the data may thus
contain mistakes. However, for a random sample of 50 papers, all
information provided on the Wiki appears to be correct.

The ReplicationWiki also includes reproductions and reinvestigations in
their definition of ``replication'', presumably because they can be
equally interesting to those evaluating the trustworthiness of a study.
For the sake of brevity, the same broad definition of ``replication'' is
used throughout this paper.

Replication pages specify whether the attempt was successful, although
this field is missing for 57\% of the studies. This information will be
used to train a classification model to distinguish between failed and
successful replication attempts. Taking this idea even further, I will
attempt to create a model that predicts from a paper's text content
whether it is likely to replicate.

Determining which paper a replication study replicates is outside the
scope of this exploratory paper, but the approach described below can be
extended to include relational information.

\hypertarget{methods}{%
\section{Methods}\label{methods}}

Replication metadata was scraped from the ReplicationWiki, including the
title of the original paper, the title of the replication paper, whether
the replication attempt was successful, whether different data was used
for the replication, and whether new methods were used. The Crossref API
was called to retrieve the DOI's of both the original papers and the
replication studies. Using the DOI's, many of the papers could be
downloaded directly. The other papers were downloaded manually. All
papers were converted to text using the command-line tool
\texttt{pdftotext}. Documents with less than 100 English words were
discarded, leaving 334 replications and 344 original papers\footnote{There
  are slightly more original papers than replications because some
  replication studies investigate multiple papers}.

Fifty-dimensional word vectors trained on Wikipedia 2014 and Gigaword 5
were obtained from the
\href{https://nlp.stanford.edu/projects/glove/}{GloVe project page}
(Pennington, Socher, and Manning 2014). The low dimensionality was
chosen because of the relatively small number of papers in the dataset.
All words in all documents were mapped to their corresponding word
vectors. Then, for each document, the average of all word vectors in
that document was used as the document representation. Document-level
hand-picked features were added on a per-task basis.

The model used in all tasks is a regularized logistic regression model
(C = 1.0, class weights were balanced using the method devised by King
and Zeng (2001)).

\hypertarget{task-1-identifying-replications}{%
\subsection{Task 1: Identifying
Replications}\label{task-1-identifying-replications}}

\emph{In this task, the goal was to identify replication studies in a
mixed set of replication studies and replicated studies.}

Replication studies and replicated papers were annotated with positive
and negative labels, respectively. For this task, the normalized
frequencies of words starting with ``replicat'', ``reproduc'', ``note'',
``comment'', ``reply'', ``re-'' and ``reinvestigat'' were independently
added as features. Additionally, both the average word vectors and the
hand-picked features were recomputed for the paper titles rather than
the full text, and this second feature matrix was concatenated to the
first one. A logistic regression model was trained and tested on a
stratified 40-fold split of the data.

\hypertarget{task-2-categorizing-replications}{%
\subsection{Task 2: Categorizing
Replications}\label{task-2-categorizing-replications}}

\emph{In this task, the goal was to distinguish failed replication
studies from successful replication studies.}

Replication studies were annotated with a label denoting whether the
replication failed. Partially successful and ambiguous replications were
discarded, as were replications where data about the outcome was
missing, leaving 150 replications, of which 49 were successful. A
logistic regression model was trained and tested on a stratified 20-fold
split of the data.

\hypertarget{task-3-predicting-failure-to-replicate}{%
\subsection{Task 3: Predicting Failure to
Replicate}\label{task-3-predicting-failure-to-replicate}}

\emph{In this task, the goal was to distinguish replicated studies with
successful replications from replicated studies with failed
replications.}

All original papers were annotated with a label denoting whether its
replications failed. Papers with mixed or partially successful
replications were discarded, as were original papers where all
replication results were missing, leaving 178 original papers, of which
68 were successfully replicated. A logistic regression model was trained
and tested on a stratified 20-fold split of the data.

\hypertarget{results}{%
\section{Results}\label{results}}

\includegraphics[width=0.33\textwidth,height=\textheight]{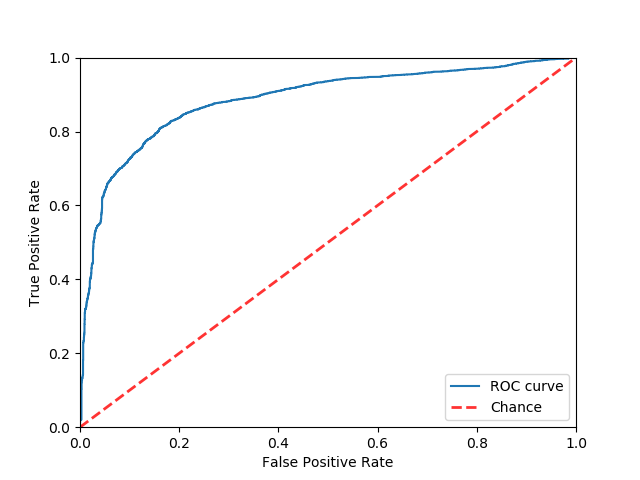}
\includegraphics[width=0.33\textwidth,height=\textheight]{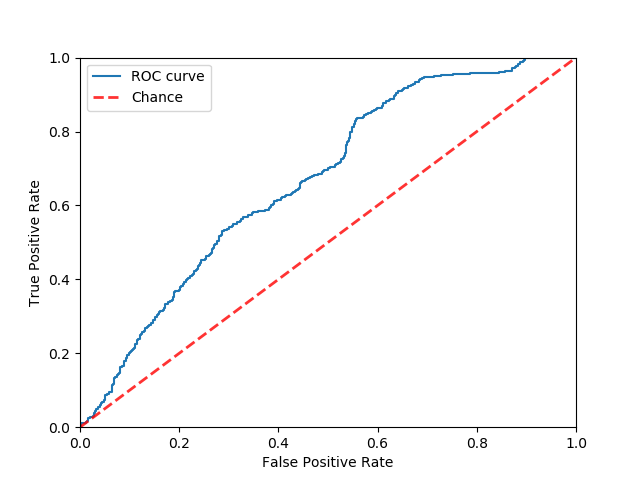}
\includegraphics[width=0.33\textwidth,height=\textheight]{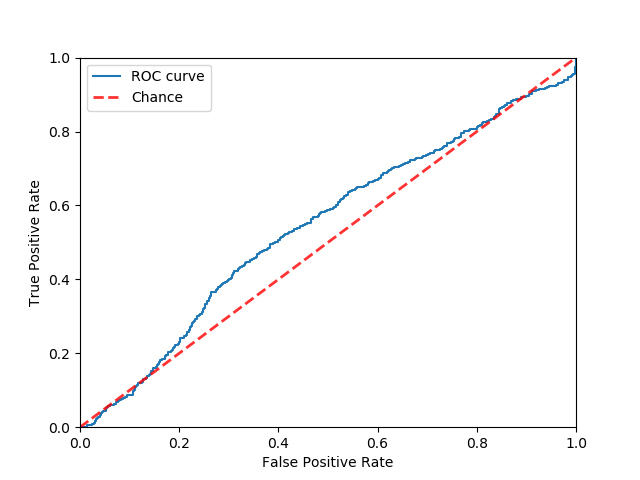}

\emph{Figure 1: ROC curves for the three tasks: identification (AUC =
0.886), categorization (AUC = 0.664), and prediction (AUC = 0.543). Red
lines indicate chance.}

\includegraphics[width=0.33\textwidth,height=\textheight]{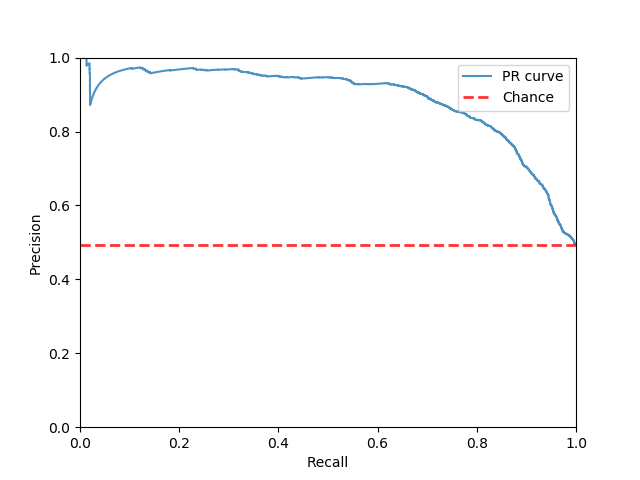}
\includegraphics[width=0.33\textwidth,height=\textheight]{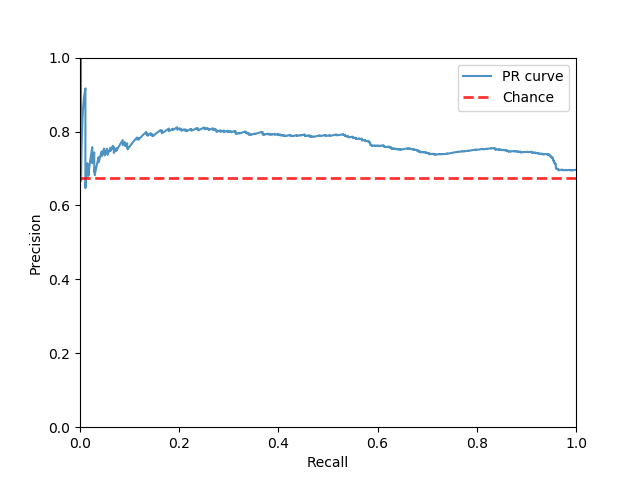}
\includegraphics[width=0.33\textwidth,height=\textheight]{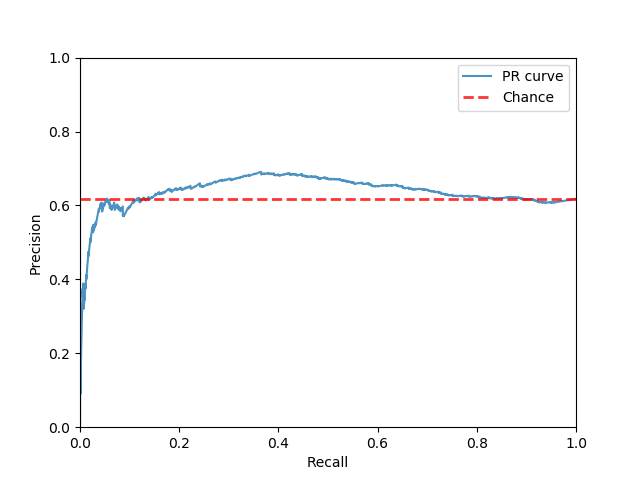}

\emph{Figure 2: Precision-recall curves for the three tasks:
identification (AUC = 0.890, chance = 0.493), categorization (AUC =
0.76, chance = 0.67), and prediction (AUC = 0.637, chance = 0.618). Red
lines indicate chance.}

As reasonably expected, predicting whether a paper fails to replicate
based on its contents is the most challenging task, followed by
categorizing failed and successful replication papers. The highest
performance is achieved on identification of replication papers.

\hypertarget{discussion}{%
\section{Discussion}\label{discussion}}

Performance on all tasks is lower than necessary for most fully
automated use cases. The small size of the dataset is likely one of the
main culprits. Future work could gather data from sources other than the
ReplicationWiki.

To keep the design simple, replicated papers were used as the
non-replication papers the replication studies had to be distinguished
from in the first task. In most use cases, however, replication papers
have to be distinguished from other papers that cite the replicated
paper. Although replicated papers may not be entirely similar to the set
of papers that cite replicated papers, the results on the identification
task at the very least showed that replication studies are
distinguishable from other papers.

Since the ReplicationWiki was initially constructed by students, there
may be some selection bias present. For example, students may opt to
only include replications they consider interesting in the database. As
such, future work could look at evaluating the classifiers on
specialized, unbiased test sets constructed for specific use cases.

\hypertarget{bibliography}{%
\section*{Bibliography}\label{bibliography}}
\addcontentsline{toc}{section}{Bibliography}

\hypertarget{refs}{}
\begin{CSLReferences}{1}{0}
\leavevmode\vadjust pre{\hypertarget{ref-2017replicationwiki}{}}%
Höffler, Jan H. 2017. {``ReplicationWiki: Improving Transparency in
Social Sciences Research.''} \emph{The Idealis}.

\leavevmode\vadjust pre{\hypertarget{ref-king2001logistic}{}}%
King, Gary, and Langche Zeng. 2001. {``Logistic Regression in Rare
Events Data.''} \emph{Political Analysis} 9 (2): 137--63.

\leavevmode\vadjust pre{\hypertarget{ref-reprod}{}}%
Nosek, Brian A, Johanna Cohoon, Mallory Kidwell, and Jeffrey R Spies.
2018. {``Estimating the Reproducibility of Psychological Science.''}
OSF. \href{https://osf.io/ezum7}{osf.io/ezum7}.

\leavevmode\vadjust pre{\hypertarget{ref-pennington2014glove}{}}%
Pennington, Jeffrey, Richard Socher, and Christopher Manning. 2014.
{``Glove: Global Vectors for Word Representation.''} In
\emph{Proceedings of the 2014 Conference on Empirical Methods in Natural
Language Processing (EMNLP)}, 1532--43.

\end{CSLReferences}

\end{document}